\documentclass[conference,letterpaper]{IEEEtran}
\usepackage{amsmath,amsthm,amssymb}
\usepackage[english]{babel}
\usepackage{graphicx}
\usepackage[nobreak,noadjust]{cite}

\let\labelindent\relax
\usepackage[hidelinks]{hyperref}
\usepackage{enumitem}
\setlist[itemize]{leftmargin=*}
\usepackage{fancyhdr}
\begin{document}

\title{Filtered-OFDM --- Enabler for Flexible Waveform in The 5th Generation Cellular Networks}

\author{Xi Zhang\IEEEauthorrefmark{1}, Ming Jia\IEEEauthorrefmark{2}, Lei Chen\IEEEauthorrefmark{1}, Jianglei Ma\IEEEauthorrefmark{2}, Jing Qiu\IEEEauthorrefmark{1}\\\IEEEauthorrefmark{1}Chengdu Research \& Development Centre, Huawei Technologies Co., Ltd., People's Republic of China\\\IEEEauthorrefmark{2}Ottawa Research \& Development Centre, Huawei Technologies Canada Co., Ltd., Canada\\Emails: \{panda.zhang, ming.jia, ray.chenlei, jianglei.ma, maggie.qiu\}@huawei.com\vspace{-2ex}}

\maketitle
\thispagestyle{fancy}

\begin{abstract}
  The underlying waveform has always been a shaping factor for each generation of the cellular networks, such as orthogonal frequency division multiplexing (OFDM) for the 4th generation cellular networks (4G). To meet the diversified and pronounced expectations upon the upcoming 5G cellular networks, here we present an enabler for flexible waveform configuration, named as filtered-OFDM (f-OFDM). With the conventional OFDM, a unified numerology is applied across the bandwidth provided, balancing among the channel characteristics and the service requirements, and the spectrum efficiency is limited by the compromise we made. In contrast, with f-OFDM, the assigned bandwidth is split up into several subbands, and different types of services are accommodated in different subbands with the most suitable waveform and numerology, leading to an improved spectrum utilization. After outlining the general framework of f-OFDM, several important design aspects are also discussed, including filter design and guard tone arrangement. In addition, an extensive comparison among the existing 5G waveform candidates is also included to illustrate the advantages of f-OFDM. Our simulations indicate that, in a specific scenario with four distinct types of services, f-OFDM provides up to $46\%$ of throughput gains over the conventional OFDM scheme.
\end{abstract}

\section{Introduction}
After years of discussions across the industry and academia, the requirements and expectations for the 5th generation (5G) cellular networks have been made clear~\cite{Chin2014,Andrews2014}. Whilst the millimeter wave is expected to deliver short-range high-speed radio access by tens of Gbps~\cite{Andrews2014,Bai2014}, the lower frequency bands (e.g., those are currently used by the 4G long-term evolution (LTE) networks) will continue to provide ubiquitous and reliable radio access, but with an improved spectrum efficiency. To this end, the air interface, especially the underlying waveform, should be revisited~\cite{Sahin2014,Banelli2014}.

In 4G LTE networks, orthogonal frequency division multiplexing (OFDM) has served as an elegant solution to combat the frequency selectivity and to boost the spectrum efficiency~\cite{3GPP2015A}. Recently, it is becoming a consensus that the basic waveform of 5G should be able to offer including but not limited to: 1. Tailored services to different needs and channel characteristics, 2. Reduced out-of-band emission (OOBE), 3. Extra tolerance to time-frequency misalignment~\cite{Sahin2014,Banelli2014}. In terms of these new requirements, OFDM appears insufficient:
\begin{enumerate}
  \item The recent development on information society has presented many new types of communication services with diversified performance requirements. For instance, to avoid collision among fast-moving vehicles, the design of vehicle-to-vehicle communication should be aiming at ultra low latency and ultra high reliability~\cite{Santa2008}. In this case, the OFDM numerology and frame structure of 4G LTE, chosen mainly for mobile broadband (MBB) service, which is not that sensitive to latency or reliability, seems not the best choice. Meanwhile, to provide sufficient coverage with low power consumption thus enabling internet-of-things~\cite{Chin2014}, instead of OFDM, a narrow-band single-carrier waveform could be preferred. Generally speaking, requiring unified numerology across the assigned bandwidth, it is difficult for OFDM to suit the needs of different types of services and the associated channel characteristics simultaneously.
  \item Despite the fact that OFDM provides a high spectrum efficiency through orthogonal frequency multiplexing, the OOBE of OFDM is still not very satisfactory. To be specific, in 4G LTE, $10\%$ of the allocated bandwidth was reserved as guard band to give space for the signals to attenuate and thus to meet the spectrum mask. Indubitably, this has been a considerable waste of the frequency resource, which is becoming ever more precious.
  \item With OFDM, the time and frequency resources are uniformly split up into many equal-sized elements to carry information. To achieve orthogonality and thus avoid inter-symbol/channel interference, stringent time and frequency alignment is required, resulting in heavy signalling for synchronization, especially for uplink transmission. Failures to establish near-perfect time-frequency alignment will lead to significant performance degradations. That is to say, OFDM requires global synchronization which comes at the price of extra signalling.
\end{enumerate}

\begin{figure*}[t]
  \centering
  \includegraphics[width=0.9\linewidth]{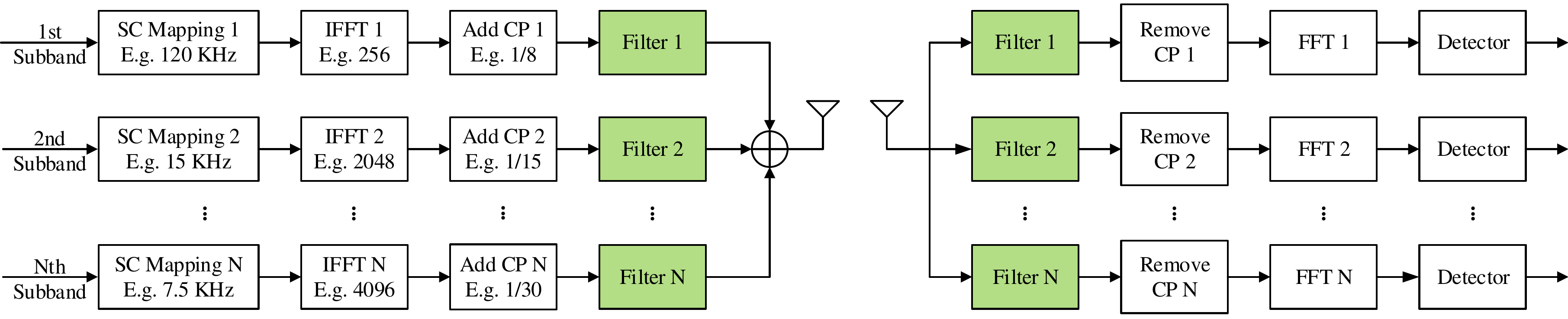}
  \caption{Downlink transceiver structure of f-OFDM.}
  \label{fig:F_OFDM_Transceiver_Structure}
  \vspace{-3ex}
\end{figure*}

To avoid the above-mentioned limitations of OFDM and to meet the new challenges faced by 5G waveform, here, in this paper, we present a new enabler for flexible waveform, named as filtered-OFDM (f-OFDM). With subband-based splitting and filtering, independent OFDM systems (and possibly other waveforms) are closely contained in the assigned bandwidth. In this way, f-OFDM is capable of overcoming the drawbacks of OFDM whilst retaining the advantages of it. First of all, with subband-based filtering, the requirement on global synchronization is relaxed and inter-subband asynchronous transmission can be supported. Secondly, with suitably designed filters to suppress the OOBE, the guard band consumption can be reduced to a minimum level. Thirdly, within each subband, optimized numerology can be applied to suit the needs of certain type of services. In general, the new performance requirements faced by 5G waveform can be fulfilled by f-OFDM and the overall spectrum efficiency can be improved. Among all the 5G waveform candidates~\cite{Sahin2014,Banelli2014}, to the authors at least, f-OFDM appears as the most promising one, in terms of the overall performance, the associated complexity, and the cost and
smoothness on the evolution path from 4G LTE.

\begin{figure}[b]
  \vspace{1ex}
  \centering
  \includegraphics[width=0.85\linewidth]{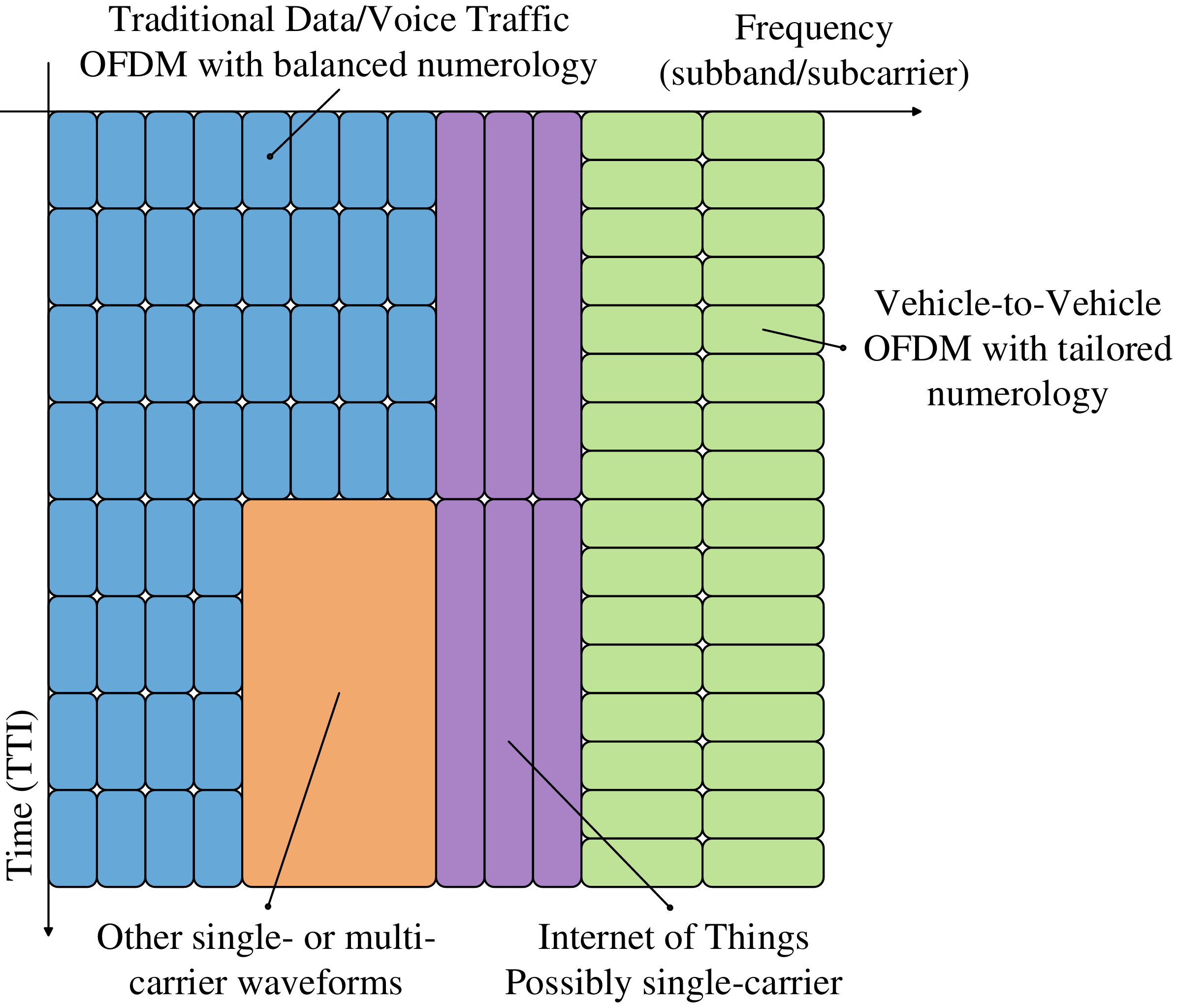}
  \caption{Flexibility and coexistence of waveforms enabled by f-OFDM.}
  \label{fig:F_OFDM_Lattice}
\end{figure}

\section{General Framework}\label{sec:General_Framework}
To boost the data rate, it is anticipated that a larger bandwidth will be allocated to 5G, e.g., 100 - 200 MHz. With f-OFDM, the assigned bandwidth will be split up into several subbands. In each subband, a conventional OFDM (and possibly other waveform) is tailored to suit the needs of certain type of service and the associated channel characteristics, e.g., with suitable subcarrier (SC) spacing, length of cyclic prefix~(CP), and transmission time interval (TTI)~\cite{3GPP2015A}, etc. Subband-based filtering is then applied to suppress the inter-subband interference, and the time-domain orthogonality between consecutive OFDM symbols in each subband is broken intentionally for a lower OOBE with negligible performance loss in other aspects. Consequently, asynchronous transmission across subbands can now be supported and global synchronization is no longer required, as opposed to the conventional OFDM. In addition, f-OFDM also provides significant reductions on the guard band consumption, leading to a more efficient spectrum utilization.

\subsection{Transceiver Structure}
The transceiver structure of f-OFDM is depicted in Fig.~\ref{fig:F_OFDM_Transceiver_Structure}. As mentioned earlier, different OFDM systems (possibly other waveforms) with different subcarrier spacing, CP length, and TTI duration are to be contained in different subbands. In general, the subbands do not overlap with each other, and between subbands, a small number of guard tones (which is much smaller than the guard band used in 4G LTE) is left to accommodate inter-subband interference and allow for asynchronous transmission. The required number of guard tones depends on the transition region of the filters and will be discussed later. The example in Fig.~\ref{fig:F_OFDM_Transceiver_Structure} is mainly for the downlink of f-OFDM. As for the uplink, subband-based filtering can also be combined with DFT-spread-OFDM~\cite{Priyanto2007}, which was used in 4G LTE networks. Here we spare this part of discussion for the sake of brevity.

\subsection{Flexibility and Coexistence}
Fig.~\ref{fig:F_OFDM_Lattice} illustrates the flexibility and coexistence of waveforms enabled by f-OFDM. As can be seen, instead of a uniform distribution as employed by OFDM in 4G LTE, the time-frequency arrangement/allocation of f-OFDM is much more flexible. For instance, to provide ultra low latency and high reliability for vehicle-to-vehicle communication~\cite{Santa2008}, the TTI duration is shortened while the subcarrier spacing of OFDM is enlarged, as compared with the OFDM numerology used in 4G LTE. Similarly, to enable sufficient coverage with low power consumption for internet-of-things~\cite{Chin2014}, a tailored single-carrier waveform is included, with possibly a small frequency occupation (thus to increase the transmit power density and overcome the penetration loss) and a long TTI duration (exploiting the quasi-static channel for transmission reliability). In general, different waveforms can be incorporated under the framework of f-OFDM, and the time-frequency arrangement may change with time, adapting to the service requirements and channel characteristics of the time.

\begin{figure}[t]
  \centering
  \includegraphics[width=0.8\linewidth]{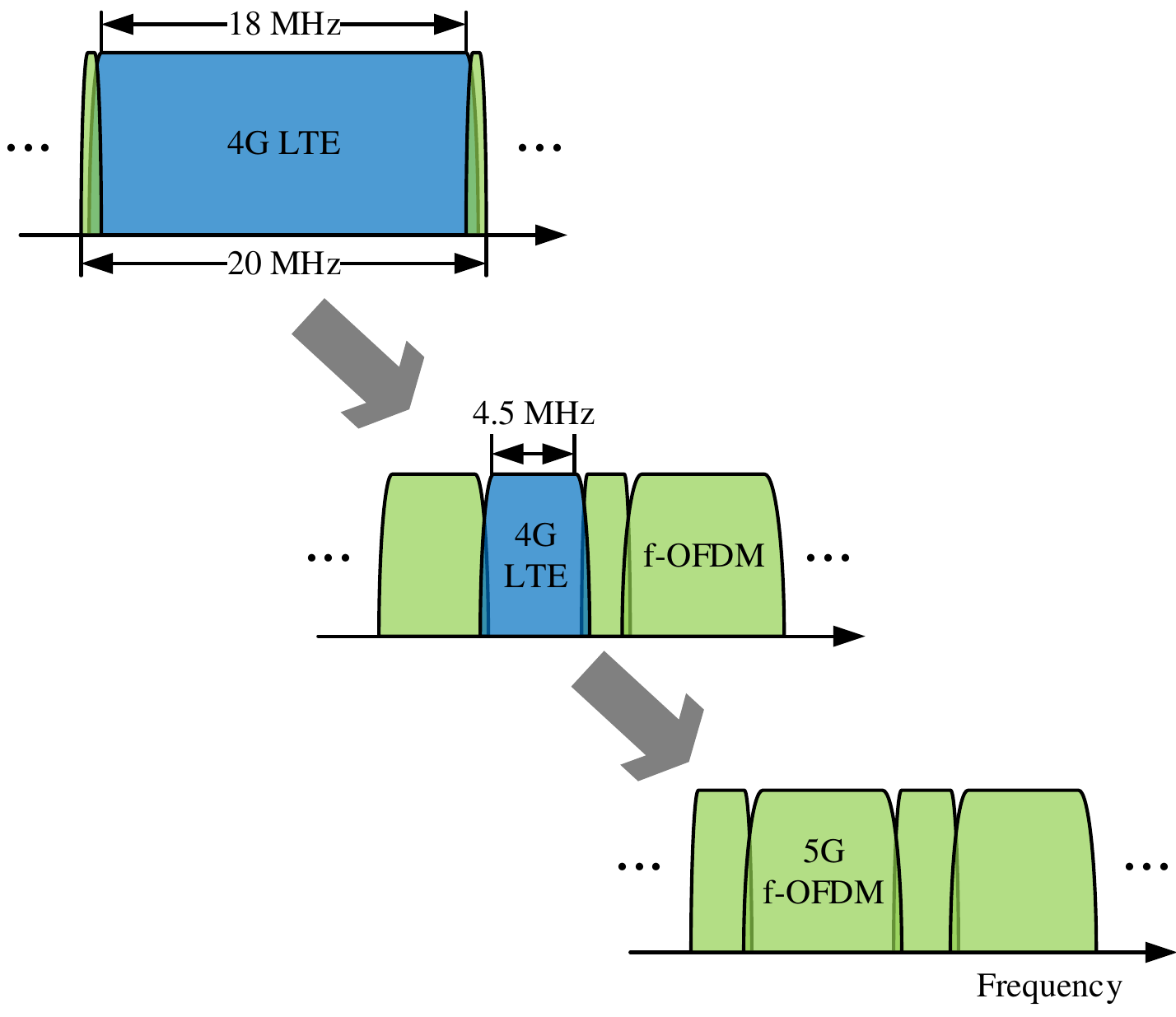}
  \caption{Possible evolution path from OFDM in 4G LTE to f-OFDM in~5G.}
  \label{fig:F_OFDM_Evolution_Path}
\end{figure}

\subsection{Evolution Path}
With independent waveforms embedded in each subband of f-OFDM, a smooth evolution from OFDM to f-OFDM can be expected. Fig.~\ref{fig:F_OFDM_Evolution_Path} presents a tentative plan for the evolution from OFDM in 4G to f-OFDM in 5G. At the initial stage, the $10\%$ guard band of the 4G networks will be invoked for data transmission through f-OFDM, and no change is required for the legacy 4G mobile devices. In the long run, the bandwidth allocated to 4G will be reduced but will likely continue to exist, while the remaining bandwidth and new spectrums will be allocated to 5G for a more flexible and efficient spectrum utilization enabled by f-OFDM. In this way, f-OFDM provides both backward and forward compatibility.

\section{Important Design Aspects}\label{sec:Important_Design_Aspects}
Here in this section, three important design aspects of f-OFDM will be discussed.

\subsection{Filter Design and Implementation}
To enable subband-based filtering and thus enjoy the benefits promised by f-OFDM, properly designed filters are needed. In general, the filter design involves the tradeoff between the time- and frequency-domain characteristics, and is also grounded by the implementation complexity. From our experience, the energy spread in the time domain shall be contained to restrain the inter-symbol interference (ISI), and the sharpness of the transition region in the frequency domain is also worth pursuing. For the sake of complexity, it is recommended to implement the filters in the frequency domain using the overlap-save method~\cite{Daher2010}, with which the benefits of fast Fourier transform (FFT) can be exploited. In addition, to achieve a flexible subband reallocation, a systematic and convenient approach that allows for online generation of filters for any given spectrum requirements is also desirable.

\begin{figure}[t]
  \centering
  \includegraphics[width=0.97\linewidth]{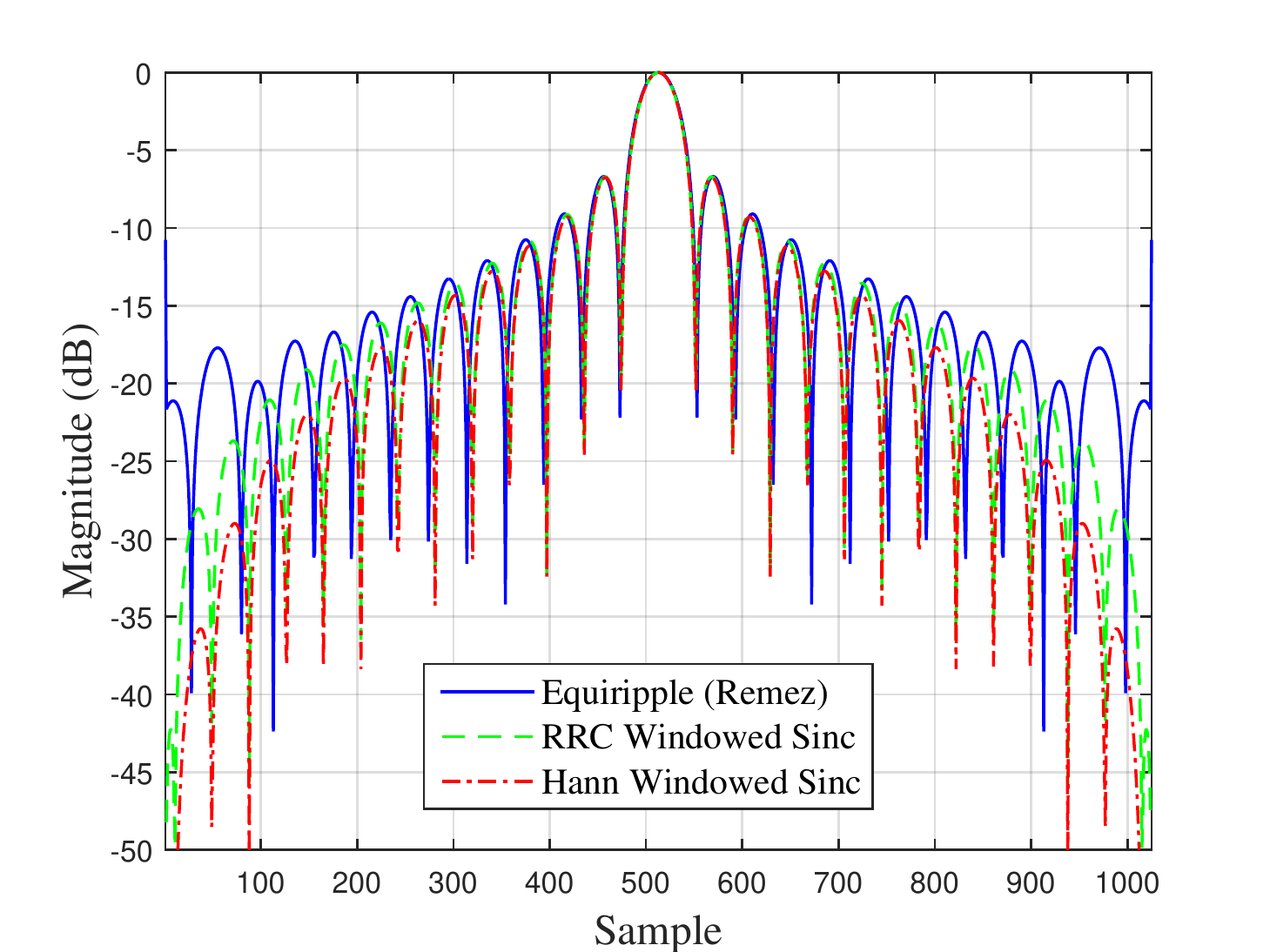}
  \caption{Impulse response of different filters.}
  \label{fig:F_OFDM_Filter_Impul_Resp}
\end{figure}

\begin{figure}[t]
  \centering
  \includegraphics[width=0.97\linewidth]{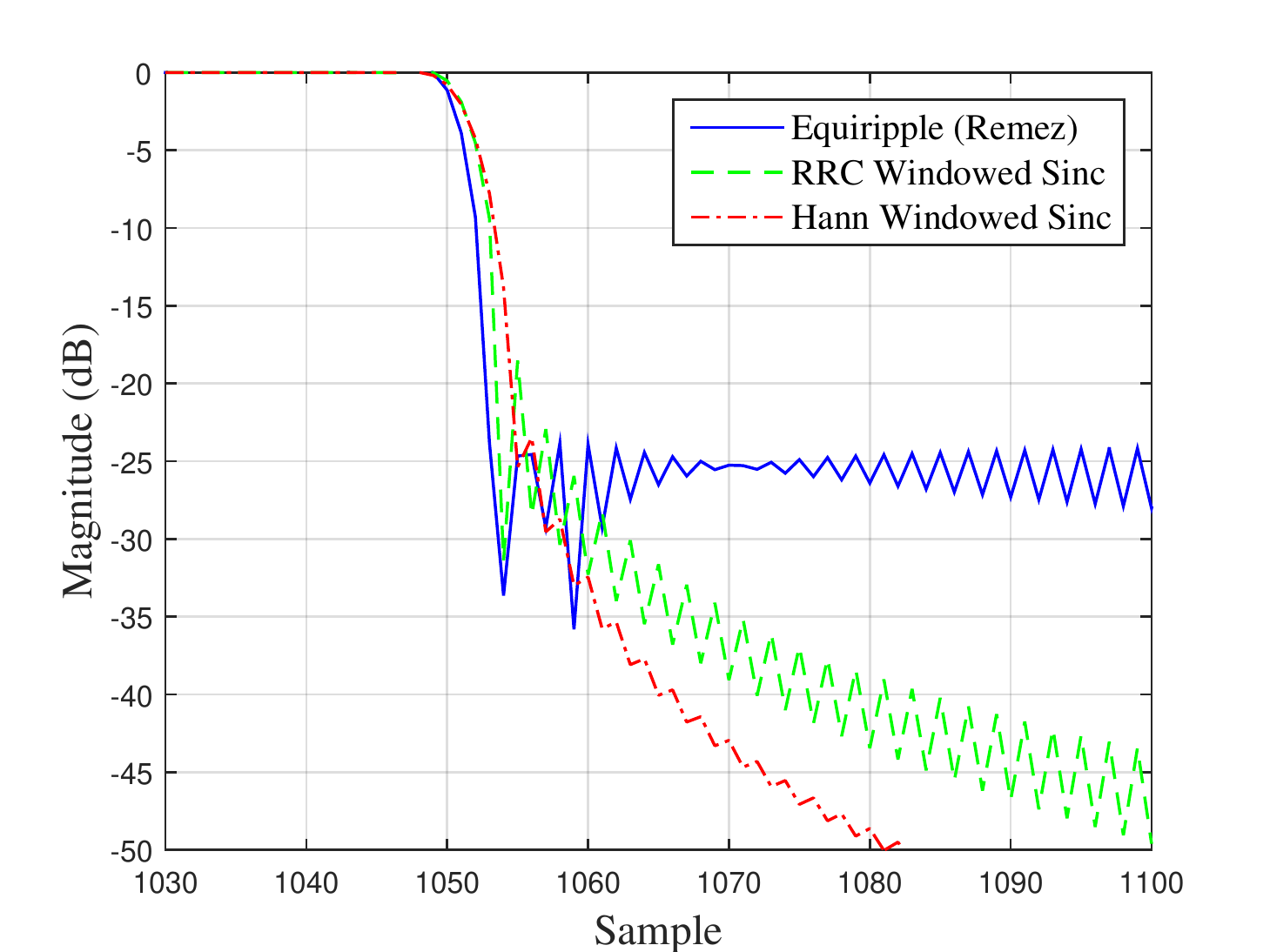}
  \caption{Frequency response of different filters.}
  \label{fig:F_OFDM_Filter_Freq_Resp}
\end{figure}

\begin{table*}[t]
  \newcommand{\tabincell}[2]{\begin{tabular}{@{}#1@{}}#2\end{tabular}}
  \renewcommand{\arraystretch}{1.25}
  \setlength{\tabcolsep}{9pt}
  \caption{Comparison among 5G Waveform Candidates}
  \centering
  \begin{tabular}{c c c c c c}
  \hline\hline
  \textbf{Waveform} & \textbf{Filter Granularity} & \textbf{Typical Filter Length} & \textbf{Time Orthogonality} & \textbf{Frequency Orthogonality} & \textbf{OOBE}\\
  \hline
  OFDM & Whole band & $\leq$ CP length & Orthogonal & Orthogonal & Bad \\ \hline
  GFDM & Subcarrier & $\gg$ Symbol duration & Non-orthogonal & Non-orthogonal & Good \\ \hline
  FBMC & Subcarrier & $=(3,4,5)~\times$ Symbol duration & Orthogonal in real domain & Orthogonal in real domain & Best \\ \hline
  UFMC & Subband & $=$ CP length & Orthogonal & Quasi-orthogonal & Good \\ \hline
  f-OFDM & Subband & $\leq1/2~\times$ Symbol duration & Non-orthogonal & Quasi-orthogonal & Better \\
  \hline\hline
  \end{tabular}
  \label{tbl:Waveform_Comparison}
  \vspace{-3ex}
\end{table*}

Two types of filters were considered in our investigation. The typical time and frequency response are depicted in Figs.~\ref{fig:F_OFDM_Filter_Impul_Resp} and~\ref{fig:F_OFDM_Filter_Freq_Resp} (with an order of 1024 and 720-KHz passband). A brief discussion about these filters are given as follows:
\begin{enumerate}
  \item \textbf{Soft-Truncated Sinc Filters}: The impulse response of an ideal low pass filter is a sinc function, which is infinitely long. For practical implementation, the sinc function is soft-truncated with different window functions: 1. Hann window; 2. Root-raised-cosine (RRC) window. In this way, the impulse response of the obtained filters will fade out quickly (see Fig.~\ref{fig:F_OFDM_Filter_Impul_Resp}), and thus limiting the ISI introduced between consecutive OFDM symbols. While being easy to generate, the soft-truncated sinc filters will have likely less resolvable frequency taps, e.g., those with a magnitude lower than -30 dB can be excluded from fixed-point frequency-domain implementation~\cite{Daher2010} (see Fig.~\ref{fig:F_OFDM_Filter_Freq_Resp}), which might lead to extra complexity savings.
  \item \textbf{Equiripple Filters}: Designed using the Remez exchange algorithm, with equiripple filters, the maximum error between the desired and the actual frequency response is minimized, and thus a sharper transition region can be obtained, as compared with the soft-truncated filters (see Fig.~\ref{fig:F_OFDM_Filter_Freq_Resp}), which, as discussed previously, is very desirable for alleviating the inter-subband interference. However, with extremely narrow transition region, the impulse response of the equiripple filters exhibits discontinuities at the head and tail (see Fig.~\ref{fig:F_OFDM_Filter_Impul_Resp}), which could be a performance-limiting factor when the operating signal-to-noise ratio (SNR) is relatively high and the ISI becomes the dominant issue, and this does not happen with soft-truncated sinc filters. Moreover, the Remez exchange algorithm requires iterative optimization and thus is inconvenient for online filter generation.
\end{enumerate}

While it appears as a good option to use soft-truncated sinc filters, there is still space for improvement and generalization, which requires further investigation.

\subsection{Guard Tone Arrangement}
In practice, the designed system will likely support only a finite set of subcarrier spacing, e.g., an integer or fractional copies of the standard subcarrier spacing used in 4G LTE (i.e., 15 KHz). The guard band between subbands, on the other hand, will likely be set as a plurality of the standard subcarrier spacing (i.e., $N\times15$ KHz), and this is why we named them as guard tones. With properly designed filters, the guard tones between subbands can be minimized to maximize the spectrum utilization. Our simulation results, which will be shown in Section \ref{sec:Simulation_Results}, has indicated that, with equal transmit power in adjacent subbands: 1. For low to medium modulation orders (e.g., QPSK and 16QAM), no guard tone is required between subbands; 2. For high-order modulation (e.g., 64QAM), up to two guard tones (i.e., two subcarriers in 4G LTE terminology) are needed between subbands. Even if the transmit power in adjacent subbands is lifted by 10 dB, two guard tones are still sufficient for suppressing the inter-subband interference for all the cases. Moreover, if combined with proper scheduling, e.g., putting high-order modulations away from the subband edges, the required number of guard tones can be further reduced.

\begin{figure}[b]
  \centering
  \includegraphics[width=0.65\linewidth]{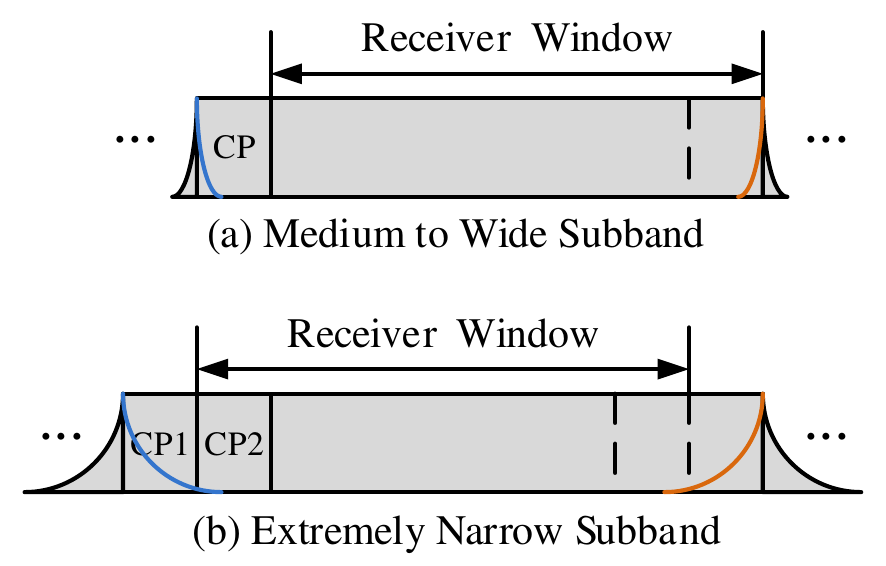}
  \caption{Treatment with filter tails in f-OFDM.}
  \label{fig:F_OFDM_Tail_Processing}
\end{figure}

\subsection{Treatment with Filter Tails}
With OFDM, the forward ISI created by multi-path can be suppressed elegantly using CP - if the length of CP exceeds the delay spread. This has been the guideline for choosing the CP duration. Now, with f-OFDM, the filters will lead to long tails in the time domain, both forward and backward. In this case, extending CP to cover the multi-path delay spread and the whole filter tails would result in a high CP overhead and thus is not a good option. Fortunately, if the subband under consideration is of a medium to large bandwidth, the mainlobe of the corresponding filter would be reasonably narrow (see Fig.~\ref{fig:F_OFDM_Filter_Impul_Resp}). Hence, in most cases, no special treatment is required for the tails that came with the filters. In extreme cases, if the passband of the filters is extremely narrow (e.g., 180 KHz), the mainlobe will spread out and the tails become non-negligible. For these cases, at the transmitter, one can extend the CP to include the main lobe of the filter and then move the receiver window forward by half of the mainlobe. In this way, both the forward and backward ISI created by filtering can be sufficiently suppressed, as indicated in Fig.~\ref{fig:F_OFDM_Tail_Processing}. However, this type of extended CP and receiver processing is needed only in the cases with extremely narrow subband, which rarely happen in practical systems.

\begin{figure}[b]
  \centering
  \includegraphics[width=0.9\linewidth]{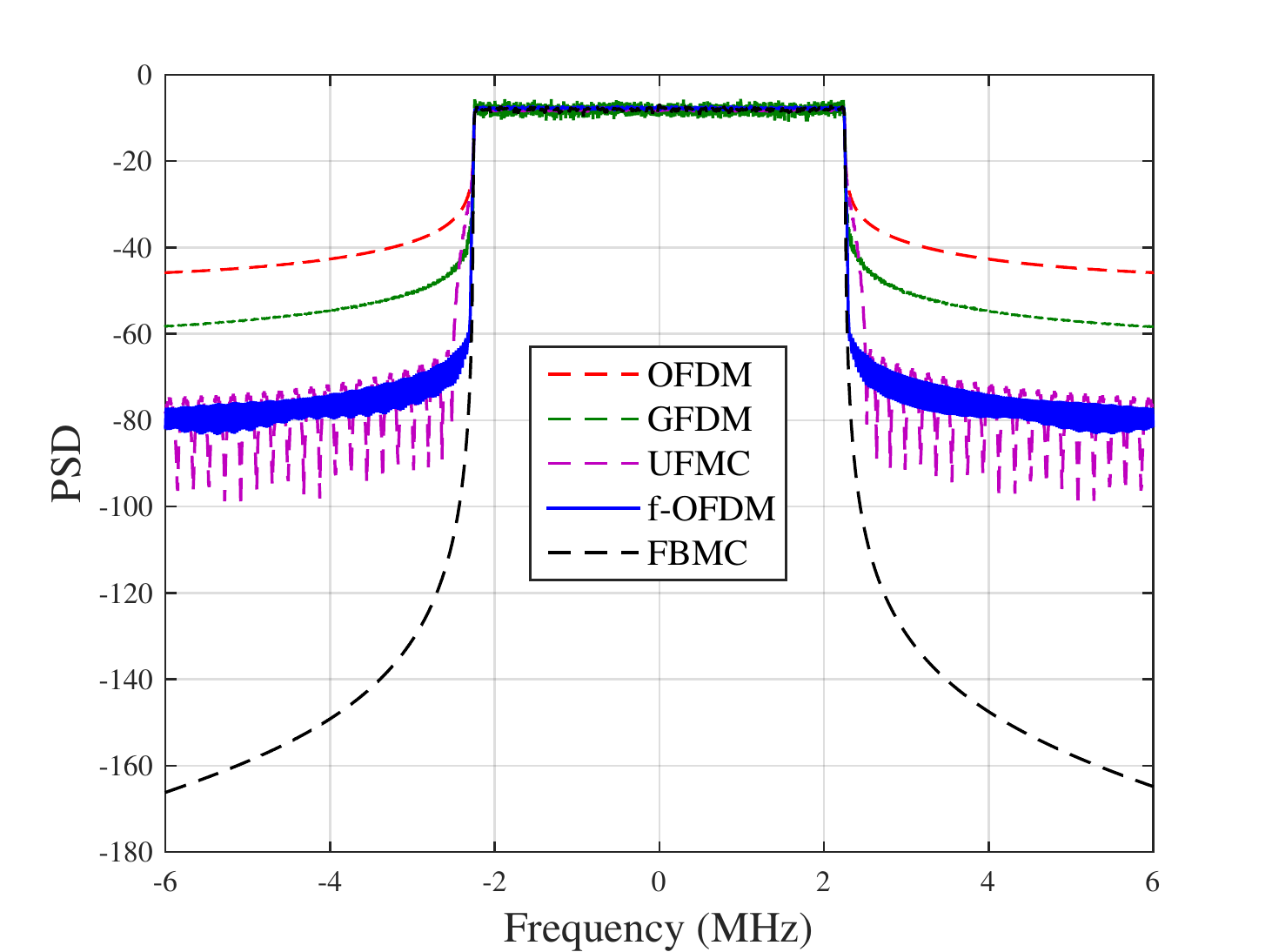}
  \caption{Typical PSD of OFDM, GFDM, UFMC, f-OFDM and FBMC, under the same level of overhead consumption.}
  \label{fig:F_OFDM_OOBE_Comparison}
\end{figure}

\section{Comparison among 5G Waveform Candidates}
Except for OFDM and f-OFDM, the most discussed 5G waveform candidates include: 1. Generalized frequency division multiplexing (GFDM)~\cite{Fettweis2009}, 2. Filter bank multi-carrier (FBMC)~\cite{Schaich2010}, 3. Universal filtered multi-carrier (UFMC)~\cite{Schaich2014B}. The motivations for these waveforms are similar to f-OFDM, i.e., to reduce OOBE or relax the requirement on synchronization. Filters are applied by all these waveforms, but with different methodology and performance. A detailed comparison among the 5G waveform candidates can be found in Table~\ref{tbl:Waveform_Comparison} and a brief discussion is given as follows.

\begin{enumerate}
  \item \textbf{f-OFDM vs. GFDM}: The subcarriers of GFDM are arranged in close proximity and are not mutually orthogonal. To suppress inter-subcarrier interference, high-order filtering and tail biting are needed~\cite{Fettweis2009}. In addition, pre-cancellation or successive interference cancellation is also required to alleviate the inter-subcarrier interference that still exists after filtering. Opposingly, the subcarriers in each subband of f-OFDM are still quasi-orthogonal, the filter length of f-OFDM is comparatively short, and no complicated pre-/post- processing is required.
  \item \textbf{f-OFDM vs. FBMC}: In pursuit of time and frequency localization, the filter length in FBMC is typically very long (e.g., more than 3 times of the symbol duration) and thus is resource-consuming, as compared with the filters in f-OFDM. Moveover, massive antenna transmission has been recognized as the cornerstone of 5G~\cite{Andrews2014}. Hence, the hardship of combining FBMC with multi-antenna transmission has limited its applications~\cite{Renfors2010}. Contrarily, f-OFDM can be combined with multi-antenna transmission without any special processing.
  \item \textbf{f-OFDM vs. UFMC}: To avoid the ISI between consecutive OFDM symbols, the filter length of UFMC is typically limited by the length of CP used in OFDM~\cite{Schaich2014B}, with which the close-in OOBE could be unsatisfactory (see Fig.~\ref{fig:F_OFDM_OOBE_Comparison}). In sharp contrast, by using a filter length up to half a symbol duration, f-OFDM intentionally gives up the orthogonality between consecutive OFDM symbols, in trade for a lower OOBE, and thus allowing for a minimum number of guard tones to be used. With properly designed filters (e.g., with a limited energy spread), the performance degradation resulted from increasing the filter length is almost negligible, as compared with the savings on guard band consumption.
\end{enumerate}

\begin{table}[b]
  \newcommand{\tabincell}[2]{\begin{tabular}{@{}#1@{}}#2\end{tabular}}
  \renewcommand{\arraystretch}{1.25}
  \setlength{\tabcolsep}{9pt}
  \caption{Simulation Setup}
  \centering
  \begin{tabular}{c c}
  \hline\hline
  \textbf{Parameter} & \textbf{Value} / \textbf{Description} \\
  \hline
  Antenna mode & $2\times2$ closed-loop beamforming \\
  Channel model & Urban macro \\
  Velocity & 3 km/h \\
  Power amplifier input backoff & 9.6 dB\\
  Type of filters & Hann windowed sinc filters \\
  Order of filters & 1024 \\
  \hline\hline
  \end{tabular}
  \label{tbl:Simulation_Setup}
\end{table}

In general, f-OFDM appears as the most promising waveform contender for 5G, providing not only the advantages of OFDM: 1. Flexible frequency multiplexing, 2. Simple channel equalization, 3. Easy combination with multi-antenna transmission, but also many new benefits: 1. Tailored services to different needs, 2. Efficient spectrum utilization, 3. Low OOBE, 4. Affordable computational complexity, 5. Possibility to incorporate other waveforms, and, 6. Backward and forward compatibility.

\section{Simulation Results}\label{sec:Simulation_Results}
Here we present two parts of simulation results. Following the 4G LTE standard, the first part includes a detailed verification of the OOBE and block error rate (BLER) of f-OFDM, corresponding to the first stage of the evolution path in Fig.~\ref{fig:F_OFDM_Evolution_Path}. In the second part, a preliminary investigation of the throughput performance of f-OFDM is conducted, providing an envision into the last stage of the evolution path in Fig.~\ref{fig:F_OFDM_Evolution_Path}.

\begin{figure}[t]
  \centering
  \includegraphics[width=0.9\linewidth]{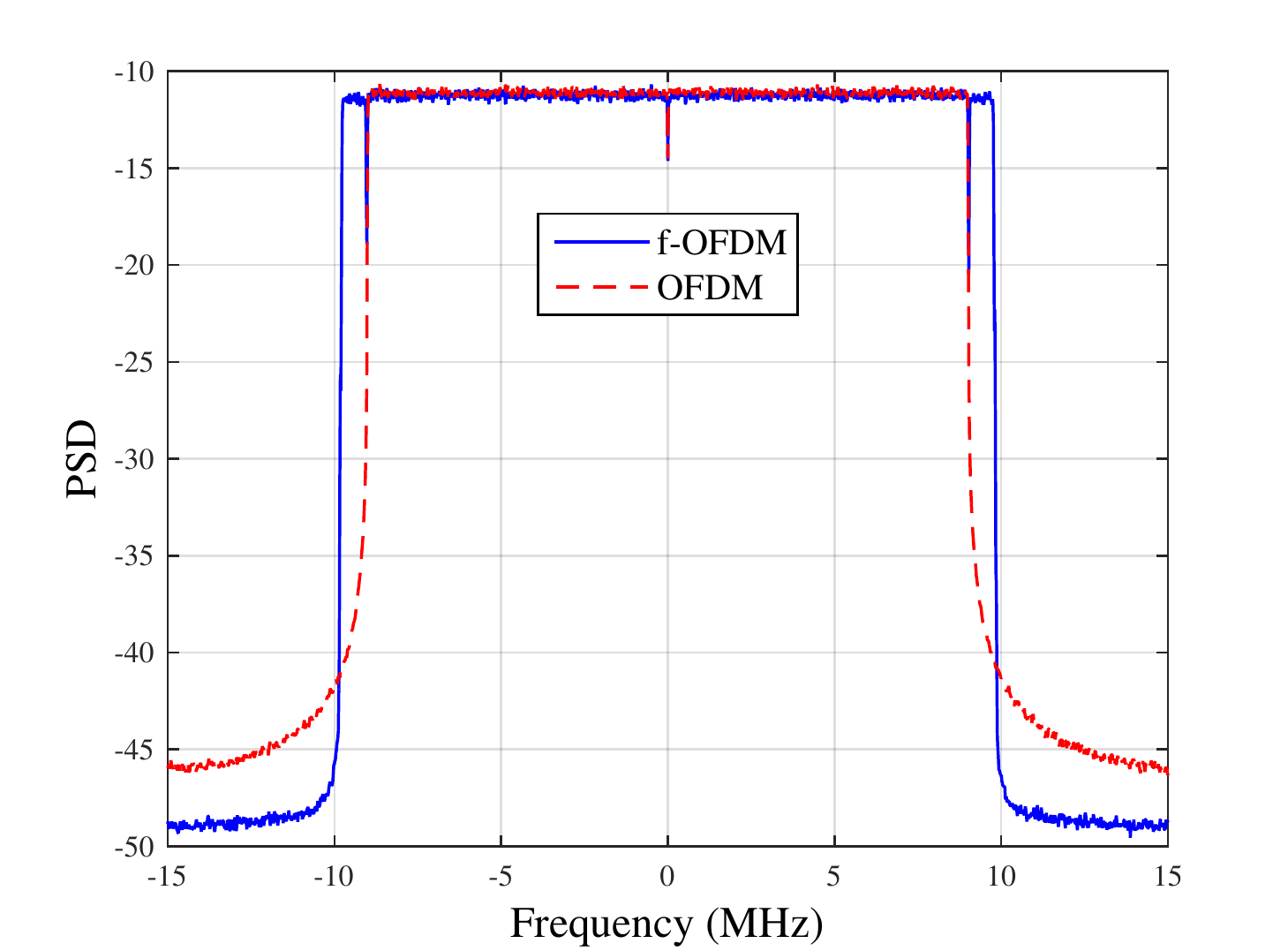}
  \caption{PSD of OFDM and f-OFDM, with distortions from power amplifier.}
  \label{fig:F_OFDM_PSD}
\end{figure}

\begin{figure}[b]
  \centering
  \includegraphics[width=0.95\linewidth]{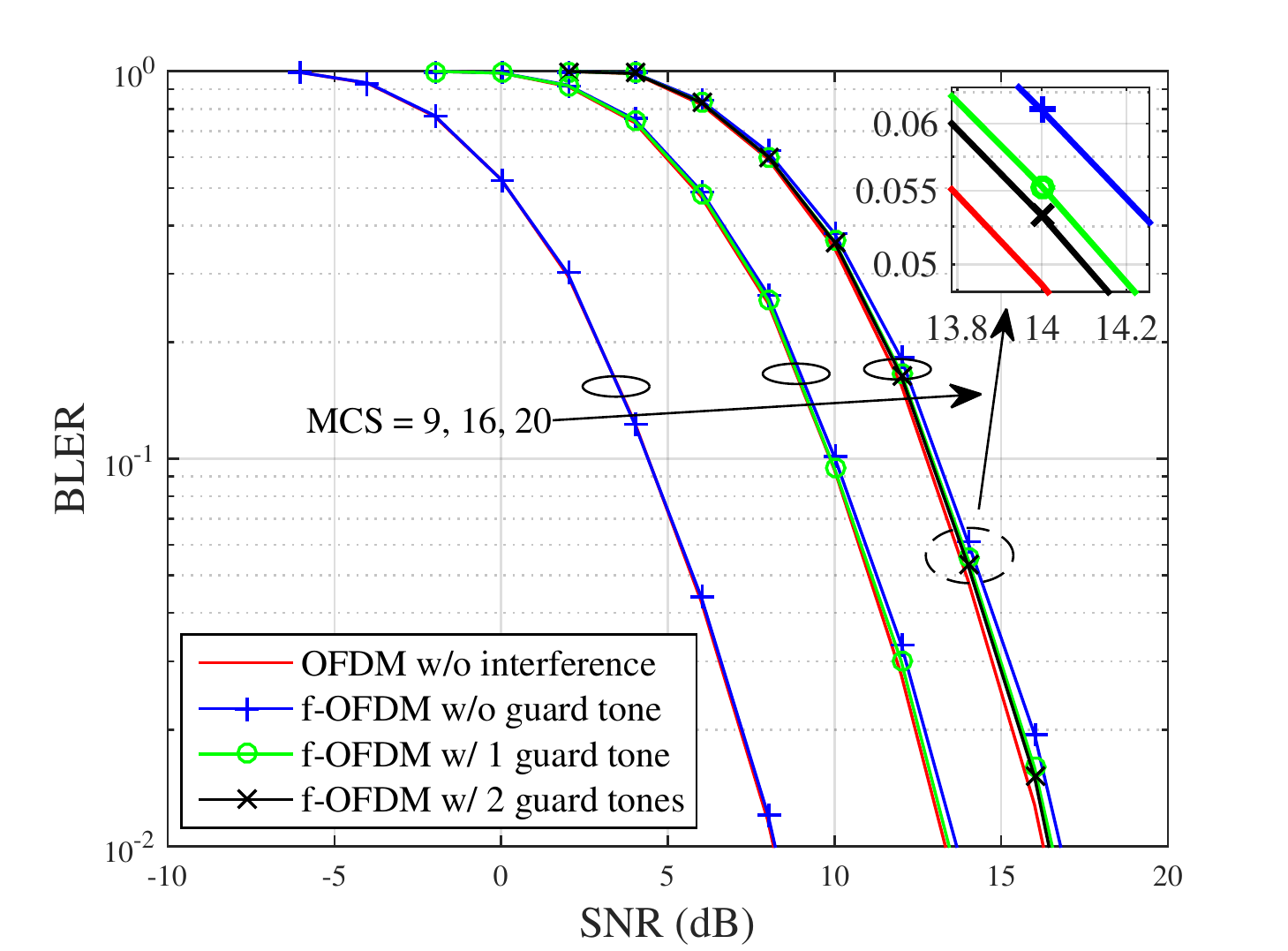}
  \caption{BLER of f-OFDM in the 1st subband with equal transmit power in all three subbands, compared with that of a single OFDM in the 1st subband without interference.}
  \label{fig:F_OFDM_BLER_1st_Subband_Guard_Tone}
\end{figure}

\begin{figure}[t]
  \centering
  \includegraphics[width=0.95\linewidth]{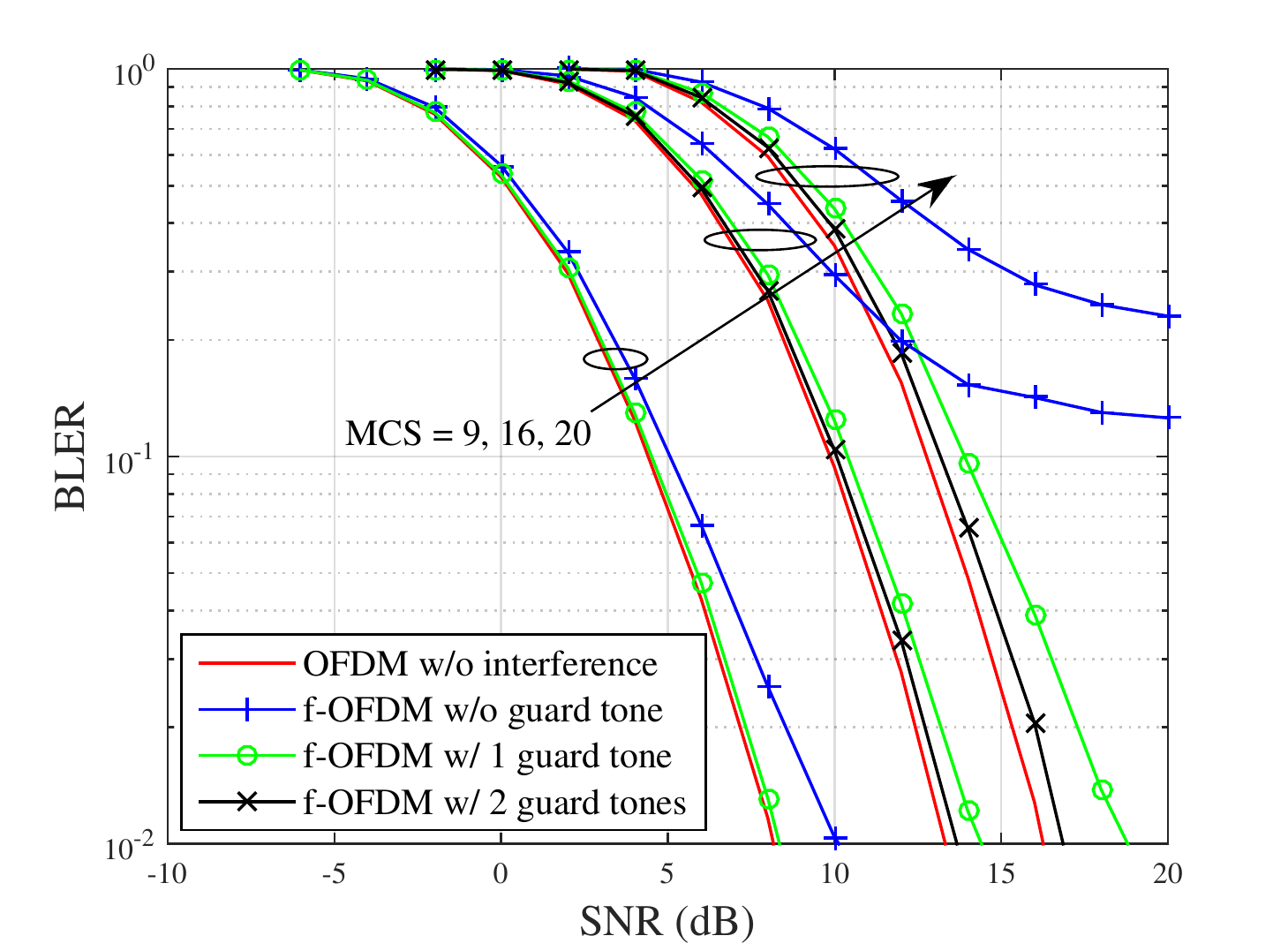}
  \caption{BLER of f-OFDM in the 1st subband while the transmit power in the 2nd subband is lifted by 10 dB, compared with that of a single OFDM in the 1st subband without interference.}
  \label{fig:F_OFDM_BLER_1st_Subband_Power_Offset}
\end{figure}

\subsection{OOBE and BLER}
In this part, the 20-MHz bandwidth of 4G LTE is divided into three subbands. The 1st and 3rd have 1-MHz bandwidth, corresponds to the guard bands in LTE. The 2nd subband has 18-MHz bandwidth, corresponds to the bandwidth loaded in LTE systems. Consistent OFDM numerology is applied across all three subbands, i.e., using uniform subcarrier spacing, normal CP and no special treatment for filter tails. The 1st and 3rd subbands are loaded with 4 resource blocks (RBs) and the 2nd is loaded with 100 RBs, with different numbers of guard tones between them. A half-symbol delay is created between adjacent subbands to simulate inter-subband interference and asynchronous transmission. The modulation and coding scheme (MCS) represents the modulation order and coding rate~\cite{3GPP2015A}. The simulation setup is summarized in Table~\ref{tbl:Simulation_Setup}, and other unlisted issues are set to follow the LTE standard. As depicted in Fig.~\ref{fig:F_OFDM_PSD}, even with the utilization of the 1-MHz guard bands, the OOBE of f-OFDM is still much lower than that of OFDM. As shown in Fig.~\ref{fig:F_OFDM_BLER_1st_Subband_Guard_Tone}, with f-OFDM, the 1st subband (i.e., the guard band in LTE systems) is efficiently utilized for data transmission with a BLER that is similar to that of a single OFDM without any interference. As demonstrated in Fig.~\ref{fig:F_OFDM_BLER_1st_Subband_Power_Offset}, even if the transmit power in the 2nd subband is lifted by 10 dB, two guard tones are still sufficient for suppressing the inter-subband interference. The BLER in the 2nd subband is similar to that of the 1st subband and thus is omitted for brevity.

\subsection{Throughput}
\begin{figure}[t]
  \centering
  \includegraphics[width=0.88\linewidth]{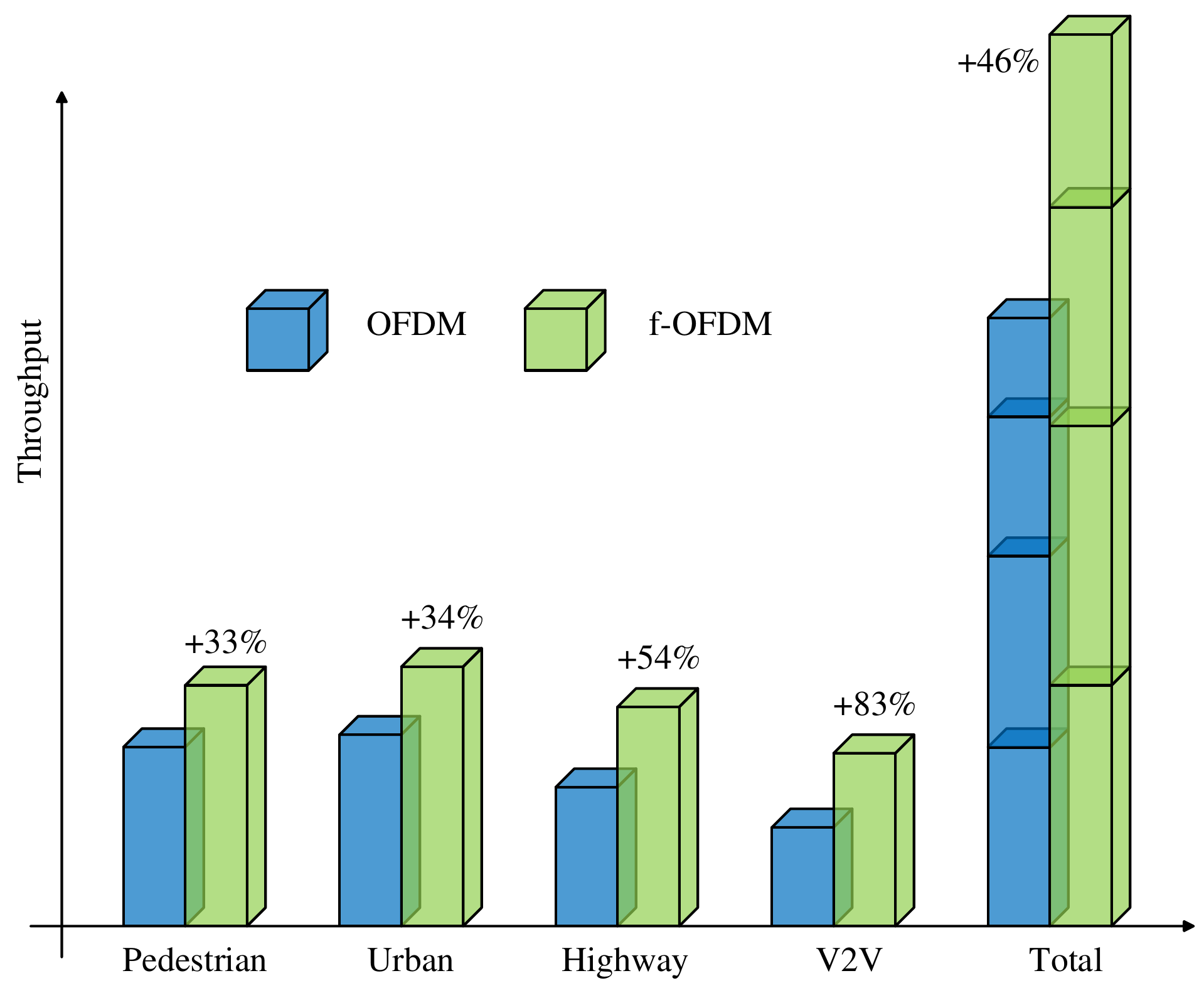}
  \caption{Normalized throughput of OFDM and f-OFDM.}
  \label{fig:F_OFDM_Throughput}
\end{figure}

Now we take an initial step to investigate the throughput gain of f-OFDM over OFDM. Our model employs single-antenna transmission. Four types of services are to be supported: 1. Pedestrian, 2. Urban, 3. Highway, 4. V2V. The numerology of OFDM is limited by the worst case, e.g., extended CP is used to combat rich multi-path scattering in urban environment, and $10\%$ bandwidth is reserved as guard band. With f-OFDM, the number of guard tones is minimized and the bandwidth available is evenly split up into four subbands. In each subband, the OFDM numerology is optimized according to the channel characteristics and service requirements (See Table~\ref{tbl:Optimized_Parameters}). As shown in Fig.~\ref{fig:F_OFDM_Throughput}, significant throughput gains can be achieved in each subband and in total. The throughput gain comes from not only the savings on guard tones, but also the adaptations to the channel characteristics, i.e., reduced CP length for smaller multi-path delay spread and reduced subcarrier spacing for stronger frequency selectivity. In this toy example, the total throughput gain of f-OFDM over OFDM is up to $46\%$, which is very attractive.

\section{Conclusions and Future Work}\label{sec:Conclusions_and_Future_Work}
In this paper, we presented filtered-OFDM --- an enabler for flexible waveform, designed to meet the expectations upon the 5G cellular networks. After outlining the general framework and methodology of f-OFDM, a detailed comparison among the 5G waveform candidates was provided to illustrate the advantages of f-OFDM. Encouraging results were observed in simulations, and to the authors at least, f-OFDM appears as the most promising 5G waveform candidate. Prototyping and field testing of f-OFDM are now in progress.

\begin{table}[t]
  \newcommand{\tabincell}[2]{\begin{tabular}{@{}#1@{}}#2\end{tabular}}
  \renewcommand{\arraystretch}{1.25}
  \setlength{\tabcolsep}{1pt}
  \caption{Optimized OFDM Numerology for Different Scenarios}
  \centering
  \begin{tabular}{c c c c c}
  \hline\hline
   & \textbf{Pedestrian} & \textbf{Urban} & \textbf{Highway}& \textbf{V2V} \\
  \hline
  Channel model & EPA~\cite{3GPP2015B} & ETU~\cite{3GPP2015B} & EVA~\cite{3GPP2015B} & Modified EVA \\
  Velocity (km/h) & 3 & 1 & 120 & 240 \\ \hline
  Subcarrier spacing (KHz) & 3.75 & 3.75 & 30 & 60 \\
  Symbol duration ($\mu s$) & 266.67 & 266.67 & 33.33 & 16.67 \\
  CP length ($\mu s$) & 2.6 & 7.49 & 2.93 & 1.95 \\
  \hline\hline
  \end{tabular}
  \label{tbl:Optimized_Parameters}
\end{table}

\bibliographystyle{IEEEtran}
\bibliography{Cited}

\end{document}